\documentstyle[prc,twocolumn,aps]{revtex}

\input{BoxedEPS.tex}
\SetRokickiEPSFSpecial
\HideDisplacementBoxes

\def\lsim{\mathrel{\rlap{\lower4pt\hbox{\hskip1pt$\sim$}}
    \raise1pt\hbox{$<$}}}         
\def\gsim{\mathrel {\rlap{\lower4pt\hbox{\hskip1pt$\sim$}}
    \raise1pt\hbox{$>$}}}         

\preprint{NT@UW-xx-xx}
\begin{document}
\title{Electron-deuteron scattering in a current-conserving
  description of relativistic bound states: including
  meson-exchange-current contributions}
\author{D.~R.~Phillips$^{a,b}$, S.~J.~Wallace$^{b}$,
  N.~K.~Devine$^{c}$ } \address{ $^{a}$Department of Physics,
  University of Washington, Box 351560, Seattle, Washington 98195-1560\\ 
  $^{b}$Department of Physics, University of Maryland, College Park,
  Maryland 20742-4111 \\ $^{c}$ General Sciences Corporation, 
  4600 Powder Mill Rd., Suite 400, Beltsville, Maryland 20705-1675}
\maketitle \narrowtext
\begin{abstract}
  Using a three-dimensional formalism that includes relativistic
  kinematics, the effects of negative-energy states, approximate
  boosts of the two-body system, and current conservation we calculate
  the electromagnetic form factors of the deuteron up to $Q^2=6$
  GeV$^2$. This is done both in impulse approximation and with a $\rho
  \pi \gamma$ meson-exchange current included. The
  experimentally-measured quantities $A$, $B$, and $T_{20}$ are
  calculated over the kinematic range probed in recent Jefferson
  Laboratory experiments. The meson-exchange current provides
  significant strength in $A$ at large $Q^2$, but has
  little impact on $B$ or $T_{20}$.
\end{abstract}
\pacs{}

\vspace{-0.8cm}
\narrowtext

Theoretically the presence of a D-state in the deuteron means that
three independent form factors can be constructed: the charge,
magnetic, and quadrupole deuteron form factors, $F_C$, $F_M$, and
$F_Q$. These are related to the Breit frame matrix elements of the
deuteron electromagnetic current operator ${\cal A}_\mu$ 
in the three deuteron
magnetic sub-states $|+1 \rangle$, $|0 \rangle$, and $|-1 \rangle$ 
via the formulae: 

\begin{eqnarray}
F_C&=&\frac{1}{3\sqrt{1 + \eta}e} (\langle 0|{\cal A}_0| 0 \rangle 
+ 2 \langle +1|{\cal A}_0|+1 \rangle), \label{eq:FC}\\
F_Q&=&\frac{1}{2 \eta \sqrt{1 + \eta} e} (\langle 0|{\cal A}_0| 0 \rangle
- \langle +1|{\cal A}_0|+1 \rangle),\\
F_M&=& \frac{-1}{\sqrt{2 \eta (1 + \eta)}e} \langle +1|{\cal A}_+|0\rangle,
\label{eq:FM}
\end{eqnarray}
with $\eta=-Q^2/(4 M_d^2)$, and $Q^2$ the square of the four-momentum
transfer to the deuteron.  Hence three experimental quantities are
required to disentangle the full electromagnetic structure of this
nucleus. Two of these---the structure functions $A$ and $B$---can be
obtained from the electron-deuteron differential cross-section using
the usual Rosenbluth separation. These are related to $F_C$, $F_Q$,
and $F_M$, as follows:

\begin{eqnarray}
A&=&F_C^2 + \frac{8}{9} \eta^2 F_Q^2 + \frac{2}{3} \eta F_M^2,\\
B&=&\frac{4}{3} \eta(1 + \eta) F_M^2.
\end{eqnarray}
The third observable of choice is $T_{20}$, the tensor-polarization
observable for electron-deuteron scattering. To obtain $T_{20}$
electrons are scattered from a polarized deuteron target. $T_{20}$ is
the ratio of a combination of differential
cross-sections for electron-deuteron scattering in the three deuteron
magnetic sub-states to
the unpolarized cross-section.  It is related to $F_C$, $F_Q$ and
$F_M$, by:

\begin{equation}
T_{20}= -\sqrt{2} \, \frac{x(x+2)+ y/2}{1 + 2(x^2 + y)};
\end{equation}
where 

\begin{eqnarray}
x=\frac{2 \eta F_Q}{3 F_C}; \, \,
y=\frac{2 \eta}{3} \left[\frac{1}{2} + (1 + \eta)
\tan^2\left(\frac{\theta_e}{2}\right)\right] \frac{F_M^2}{F_C^2}.
\end{eqnarray}

Recent experiments at the Thomas Jefferson National Accelerator
Facility (JLab) have probed the electromagnetic form factors of the
deuteron at large space-like momentum transfers. $T_{20}$ has been
measured at $Q^2$ up to almost 2 GeV$^2$, $B$ out to about 1.3
GeV$^2$, and $A$ to $Q^2=6$ GeV$^2$. At these momentum transfers
relativistic kinematics and dynamics would appear to be a necessary
ingredient of any theoretical description. Hence there has been
considerable effort invested in the construction of relativistic
formalisms for the $NN$ bound state.  If the constituents of the bound
state are understood to be the neutron and the proton then this
approach is a logical extension of the standard nonrelativistic
treatment of the $NN$ system. Furthermore, regardless of the momentum
transfer involved, it is crucial that the consequences of
electromagnetic gauge invariance be incorporated in the calculation.
Minimally this means that the electromagnetic current of the deuteron
must be conserved. Indeed, the derivation of
Eqs.~(\ref{eq:FC})-(\ref{eq:FM}) {\it assumed} that the deuteronic
current ${\cal A}_\mu$ was conserved.

Naturally, $A$, $B$, and $T_{20}$ can be calculated using a
non-relativistic $NN$ interaction which is fit to the $NN$ scattering
data. This approach has met with a significant amount of success (see,
e.g.~Refs.~\cite{RS91,Wi95}). In this note we report on an attempt to
imitate this approach using a relativistic formalism. To this end we
construct an $NN$ interaction, place it in a relativistic scattering
equation, and fit the parameters of our interaction to $NN$
data.  The electromagnetic form factors of the deuteron
predicted by this $NN$ model are then calculated. The approach
includes relativistic kinematics, negative-energy states, boost
effects, and relativistic pieces of the electromagnetic current
explicitly at all stages of the calculation.  This three-dimensional
(3D) technique has been developed and applied in
Refs.~\cite{PW96,PW97,PW98}. Here our focus is on elastic
electron-deuteron scattering.  First, we give a brief review of our
formalism in which we display expressions for the bound-state equation
and the current matrix element in the case of an instantaneous
two-body interaction. Second, we discuss the inclusion of
meson-exchange-current (MEC) contributions, especially the $\rho \pi
\gamma$ MEC, which is known to give significant contributions to
electron-deuteron scattering. Then, we present our results for $A$,
$B$, and $T_{20}$.  While $T_{20}$ is reproduced quite well there is
significant discrepancy between our calculation and the experimental
data for $A$ and $B$.

A number of alternative 3D relativistic treatments of deuteron
dynamics exist (see, for instance,
Refs.~\cite{Ar80,HT,vO95,Ch88,CK99}). Of these, the closest to this
work is that of Hummel and Tjon~\cite{HT}, although we 
eliminate some approximations made there. Also, we do not include the
$\omega \sigma \gamma$ MEC in our calculation.  Our impulse
approximation results are very similar to those of
Ref.~\cite{HT}.

Consider the Bethe-Salpeter equation (BSE) for a bound-state vertex
function, $\Gamma$:

\begin{equation}
\Gamma=K G_0 \Gamma.
\label{eq:LBSE}
\end{equation}
Here $K$ is, in principle, the sum of all two-particle-irreducible $NN
\rightarrow NN$ graphs. The $NN$ propagator $G_0$ is the product of
spin-half Feynman propagators for each nucleon: $G_0=i d_1 d_2$.  In
studies of this equation for the deuteron bound state~\cite{FT} the
kernel $K$ included a set of single-boson exchanges---in analogy to
many non-relativistic potential models---yielding the ``ladder''
approximation.  However, it is well known that in such an
approximation the Bethe-Salpeter equation does not give the correct
one-body limit~\cite{Gr82}. In other words, if we consider
unequal-mass particles, and take one of them to be very heavy,
Eq.~(\ref{eq:LBSE}) does not become the Dirac equation for the light
particle moving in the static field of the heavy one. This limit is
only properly treated in Eq.~(\ref{eq:LBSE}) if the full set of ladder
and crossed-ladder graphs is taken for $K$~\cite{Gr82}. In
Ref.~\cite{PW96} we attempted to remedy this, and showed that the
pieces of the graphs which appear in $K$ and are responsible for the
one-body limit can be resummed so that Eq.~(\ref{eq:LBSE}) becomes:

\begin{equation}
\Gamma=U (G_0 + G_C) \Gamma,
\label{eq:4DET}
\end{equation}
where the precise form of $G_C$ was derived in \cite{PW96,PW98}.  For
exact correspondence between (\ref{eq:LBSE}) and (\ref{eq:4DET}) we
should have:

\begin{equation}
K=U + U G_C K.
\end{equation}
At the level of the one-boson-exchange interaction, where $K$ and $U$
have only their lowest-order pieces, we see that Eq.~(\ref{eq:4DET})
defines an improved ``ladder'' Bethe-Salpeter equation, which {\it does}
have the correct one-body limit:

\begin{equation}
\Gamma=K^{(2)} (G_0 + G_C) \Gamma.
\label{eq:LBSEplus}
\end{equation}

This equation is still four-dimensional. One straightforward way to
reduce it to three dimensions is to follow Salpeter~\cite{Sa52} and
assume an instantaneous interaction, i.e. make the replacement:

\begin{equation}
K^{(2)}(q)=\frac{1}{q^2 - \mu^2} \quad \longrightarrow \quad 
K^{(2)}_{\rm inst}({\bf q})=-\frac{1}{{\bf q}^2 + \mu^2},
\label{eq:replace}
\end{equation}
where $q=(q_0,{\bf q})$ is the four-momentum of the meson.  Since
$K^{(2)}_{\rm inst}$ depends only on the three-vector $\bf q$ this
approximation reduces Eq.~(\ref{eq:LBSEplus}) to a three-dimensional
equation:

\begin{equation}
\Gamma_{\rm inst}=K^{(2)}_{\rm inst} \langle G_0 + G_C \rangle \Gamma_{\rm inst}.
\label{eq:ET}
\end{equation}
Here the three-dimensional propagator $\langle G_0 + G_C \rangle$ is obtained 
by integrating over the time-component of relative four-momentum:

\begin{equation}
\langle G_0 + G_C \rangle=\int \frac{dp_0}{2 \pi}
\left[ G_0(p;P) + G_C(p;P) \right].
\end{equation}
Hereafter we always denote integration over zeroth components of
relative four-momenta by angled brackets.  

Before examining $\langle G_0 + G_C \rangle$, consider the more
standard equal-time Green's function~\cite{LT63,BK93B}:

\begin{equation}
\langle G_0 \rangle=\frac{\Lambda_1^+ \Lambda_2^+}{E^+ - \epsilon_1
- \epsilon_2} - \frac{\Lambda_1^- \Lambda_2^-}{E^- +  \epsilon_1
+ \epsilon_2};
\label{eq:aveG0}
\end{equation}
where $\Lambda^{\pm}$ are related to projection operators onto
positive and negative-energy states of the Dirac equation, $E$ is the
total energy, and $\epsilon_i=({\bf p}_i^2 + m_i^2)^{1/2}$. The
propagator $\langle G_0 \rangle$ is not invertible~\cite{Kl}, since it
has no components in the $+-$ and $-+$ sectors. This is related to the
lack of a correct one-body limit in the ladder BSE.  If we had applied
the 3D reduction (\ref{eq:replace}) to Eq.~(\ref{eq:LBSE}) we would
have obtained the Salpeter equation, with the non-invertible $\langle
G_0 \rangle$ in the intermediate state.  However, adding $\langle G_C
\rangle$, which comes from resumming pieces of the crossed-ladder
graphs {\it before} reducing to three dimensions, gives a 3D $NN$
propagator:

\begin{eqnarray}
  \langle G_0 + G_C \rangle&=&\frac{ \Lambda_1^+ \Lambda_2^+}{E^+ -
    \epsilon_1 - \epsilon_2} - \frac{ \Lambda_1^+ \Lambda_2^-}{2
    \kappa_2^0 - E + \epsilon_1 + \epsilon_2} \nonumber\\ &-&
  \frac{ \Lambda_1^- \Lambda_2^+}{E - 2 \kappa_2^0 + \epsilon_1
    + \epsilon_2} - \frac{ \Lambda_1^- \Lambda_2^-}{E^- + \epsilon_1 +
    \epsilon_2},
\label{eq:aveG0GCgeneral}
\end{eqnarray}
with $\kappa_2^0$ a parameter that enters through the construction of
$G_C$. This three-dimensional propagator was derived by Mandelzweig
and Wallace with $\kappa_2^0$ equal to the on-shell energy of particle
two~\cite{MW}. With $\kappa_2^0$ chosen in this way $\langle G_0 + G_C
\rangle$ has the correct one-body limits as either particle's mass
tends to infinity and also is invertible.

\begin{figure}[h,t,b]
\centerline{\BoxedEPSF{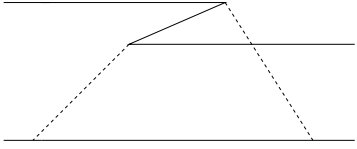 scaled 350}}

\vskip 3 mm

\caption{One example of a Z-graph which is included in our
3D equation (\ref{eq:ET}).}
\label{fig-Zgraph}  
\end{figure}

In this work we consider the interaction of two particles of equal
mass, and thus choose a different $\kappa_2^0$, namely $\kappa_2^0 =
(E - \epsilon_1 + \epsilon_2)/2$.  This form avoids the appearance of
unphysical singularities when electron-deuteron scattering is
calculated~\cite{PW98}. It yields a two-body propagator:

\begin{eqnarray}
&&  \langle G_0 + G_C \rangle=\nonumber\\
&& \quad \frac{ \Lambda_1^+ \Lambda_2^+}{E^+ -
    \epsilon_1 - \epsilon_2} - \frac{ \Lambda_1^+ \Lambda_2^-}{2
    \epsilon_2} - \frac{ \Lambda_1^- \Lambda_2^+}{2
    \epsilon_1} - \frac{ \Lambda_1^- \Lambda_2^-}{E^- +
    \epsilon_1 + \epsilon_2},
\label{eq:aveG0GC}
\end{eqnarray}
which is consistent with that demanded by low-energy theorems for
composite spin-half particles in scalar and vector fields~\cite{Ph97B}.
Alternatively, comparing the $++ \rightarrow ++$ piece of 

\begin{equation}
K^{(2)}_{\rm inst} \langle G_0 + G_C \rangle K^{(2)}_{\rm inst}
\end{equation}
with the amplitude obtained at fourth order in the full 4D field
theory we see that the contribution of negative-energy states agrees
at leading order in $1/M$~\cite{PW98}. In other words, effects such as
Fig.~\ref{fig-Zgraph} are included in a bound-state calculation that
employs Eq.~(\ref{eq:ET}). This is true even if only the instantaneous
ladder kernel $K_{\rm inst}^{(2)}$ is used, because of our careful
treatment of the one-body limit.

Of course, the instant approximation which led to Eq.~(\ref{eq:ET}) is
uncontrolled. However, one of the central results of Ref.~\cite{PW96}
was that similar three-dimensional reductions could be implemented in
a systematically-improvable way.  The numerical work of
Ref.~\cite{PW98} then showed that the replacement of
Eq.~(\ref{eq:replace}) by the systematic 3D reductions developed in
Ref.~\cite{PW96} had little impact on the deuteron electromagnetic
form factors. Hence in this work we will report only on results
obtained using Eq.~(\ref{eq:ET}), even though ``better'' treatments of
the reduction to three dimensions are certainly possible.

It remains to construct a conserved electromagnetic deuteron current.
We begin by defining ${\cal A}_\mu^{(0)}$, the impulse approximation
current obtained when the ladder Bethe-Salpeter equation
(\ref{eq:LBSE}) is solved:

\begin{eqnarray}
  {\cal A}^{(0)}_\mu&=&i \, \bar{\Gamma}(P+Q) \, d_1(p_1) \, d_2(p_2 + Q)
  j_\mu^{(2)} d_2(p_2) \, \Gamma(P)\nonumber\\ 
&& \qquad + \, \, (1 \leftrightarrow 2),
\label{eq:BSEIA}
\end{eqnarray}
where $\Gamma$ is the solution of Eq.~(\ref{eq:LBSE}), $P=p_1 + p_2$
and $P+Q$ are the four-momenta of the initial and final states, and
$j_\mu^{(n)}$ is the usual one-nucleon current for particle $n$:

\begin{equation}
  j_\mu^{(n)}=q_n \left[F_1^{(n)}(Q^2) \gamma^{(n)}_\mu + F_2^{(n)}(Q^2)
    \frac{i}{2 M} \sigma_{\mu \nu}^{(n)} Q^\nu \right]
\label{eq:job}
\end{equation}
($q_n$ is the charge). Using the Ward-Takahashi identity associated
with formally-modified but practically-identical form of this
current~\cite{GR87} it is easy to show that ${\cal A}_\mu^{(0)}$ is
conserved, i.e.

\begin{equation}
  Q^\mu {\cal A}^{(0)}_\mu=0.
\end{equation}

However, in this work we did not begin with the ladder BSE. Instead we
began with the 4D equation (\ref{eq:4DET}). Constructing a conserved
impulse approximation current for the vertex function which is the
solution of Eq.~(\ref{eq:4DET}) is a little more involved. In
Ref.~\cite{PW98} we showed how to add a piece to the current
(\ref{eq:BSEIA}) which results in ${\cal A}_\mu$ being conserved
if $\Gamma$ is the solution of Eq.~(\ref{eq:4DET}):

\begin{equation}
  {\cal A}_\mu={\cal A}^{(0)}_\mu + \bar{\Gamma}(P+Q) {G_C}_\mu \Gamma (P).
\label{eq:4DETAmu}
\end{equation}
The explicit expression for ${G_C}_\mu$ can be found in
Ref.~\cite{PW98}.

\begin{figure}[h,t,b]
\centerline{\BoxedEPSF{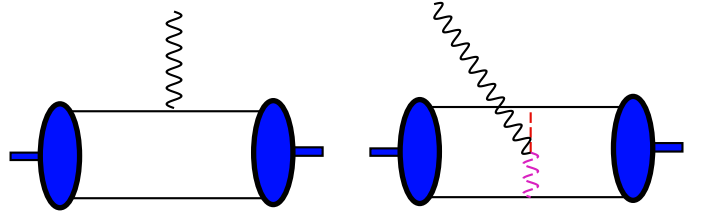 scaled 400}}

\vskip 2 mm

\caption{On the left we depict the positive-energy-state impulse approximation
  mechanism for electron-deuteron scattering. On the right the $\rho
  \pi \gamma$ MEC diagram is shown.}
\label{fig-current}  
\end{figure}

With this four-dimensional current in hand we may make a
reduction of it to three dimensions in an analogous fashion to
the reduction employed for the bound-state equation itself. Once again,
this reduction can be performed in a systematic fashion, 
but here we need only the
results for an instantaneous interaction. In that case the
``instant'' current:

\begin{equation}
{\cal A}_{{\rm inst},\mu}=\bar{\Gamma}_{\rm inst}
\, {\cal G}^\gamma_{\rm inst, \mu} \, \Gamma_{\rm inst}
\label{eq:Amuinst}
\end{equation}
is conserved, provided that $\Gamma_{\rm inst}$ is the solution of
Eq.~(\ref{eq:ET}). The explicit form of the current employed
is~\cite{PW98}:

\begin{eqnarray}
  {\cal G}_{{\rm inst},\mu}^\gamma({\bf p}_1,{\bf p}_2;P,Q)=i \,
  \langle d_1(p_1) \, d_2(p_2+Q) j^{(2)}_\mu d_2(p_2) \rangle
  \nonumber\\ + \, i \, \langle d_1(p_1) \, d_2^{\tilde{c}}(p_2+Q)
  j^{(2)}_{c,\mu} d_2^c(p_2) \rangle + (1 \leftrightarrow 2).
\end{eqnarray}
Here $d_n$ is the Dirac propagator for particle $n$, and $j_\mu^{(n)}$
is the one-body current (\ref{eq:job}). Only the $\gamma^\mu$ piece of
$j_\mu^{(n)}$ in relevant for charge conservation, since the piece
proportional to $\sigma^{\mu \nu}$ is automatically conserved.
Meanwhile, $d_n^c$ is a one-body Dirac propagator used in $G_C(P)$ to
construct the approximation to the crossed-ladder graphs.
Correspondingly, $d_n^{\tilde{c}}$ appears in $G_C(P+Q)$. It does {\it
  not} equal $d_n^c$, even if particle $n$ is not the nucleon struck
by the photon.  Finally,

\begin{equation}
j^{(2)}_{c,\mu}= (q_2 \gamma^{(2)}_\mu - \tilde{j}^{(2)}_\mu);
\quad
  \tilde{j}^{(2)}_\mu=q_2 \frac{\hat{p}_{2 \mu}' + \hat{p}_{2
      \mu}}{\epsilon_2' + \epsilon_2} \gamma^{(2)}_0,
\label{eq:tildej2}
\end{equation}
with $\hat{p}_2=(\epsilon({\bf p}_2),{\bf p}_2)$.  The current defined
by Eqs.~(\ref{eq:Amuinst})--(\ref{eq:tildej2}) includes not only the
effects of time-ordered graphs like that on the left-hand side of
Fig.~\ref{fig-current}, but also the effects of photons coupling to
the Z-graph in Fig.~\ref{fig-Zgraph}.

This defines our impulse-approximation current.  Detailed results for
this current employed with the solutions of Eq.~(\ref{eq:ET}) were
presented in Ref.~\cite{PW98}. In this work we move beyond the impulse
approximation by including the effects of the $\rho \pi \gamma$
meson-exchange current. Because of the quantum numbers of the deuteron
this is generally thought to be the lowest-mass mesonic excitation
which makes a contribution to the electromagnetic deuteron current.
However, note that there are additional pionic currents which should
be included because we have chosen pseudovector pion coupling. In
nonrelativistic approaches pseudovector pion coupling implies that
there is a ``relativistic'' two-body contribution to the deuteron's
charge operator involving a $\gamma \pi$ contact term~\cite{Ri84}.
In $A$ this effect plays a more significant role than the $\rho \pi
\gamma$ MEC~\cite{CS98}. This dynamics appears in the current operator
of the formalism derived in Ref.~\cite{PW97}, but we have not included
it in this calculation.

The Lagrangian governing the $\rho \pi \gamma$ vertex is:

\begin{equation}
  {\cal L}\, =-e \, \frac{g_{\rho \pi \gamma}}{2 m_\rho} \,
  \epsilon_{\alpha \beta \gamma \delta} F^{\alpha \beta} \,
  \vec{\rho}^{\, \gamma} \cdot \partial^\delta \vec{\pi}.
\end{equation}
This yields the two-body current depicted on the right-hand side of
Fig.~\ref{fig-current}. In calculating this diagram the same form
factors are employed at the $\pi NN$ and $\rho NN$ vertices as were
used in calculating the $NN$ potential. The $\rho \pi \gamma$ current
is automatically conserved.

With all the theoretical pieces of the puzzle assembled we now
calculate electron-deuteron scattering. The vertex functions employed
are the ones calculated with all positive and negative-energy states
included, as described in Ref.~\cite{PW98}.  If the negative-energy
states are dropped the interaction is exactly the Bonn-B potential for
the Thompson equation, as derived and fitted to $NN$ phase shifts in
Ref.~\cite{Ma89}. This model gives a reasonable fit to the $NN$ data,
and good deuteron static properties, although it is not as good a fit
as some more recent $NN$ potentials~\cite{Wi95,St94,Ma96}. When
negative-energy states are included the deuteron pole position changes
slightly. To compensate for this we adjust the $\sigma$-meson coupling
from the value of the fit in Ref.~\cite{Ma89}, $\frac{g_\sigma^2}{4
  \pi}=8.08$, to $\frac{g_\sigma^2}{4 \pi}=8.55$.

The single-nucleon form factor parametrization chosen is that of
Mergell {\it et al.}~\cite{Me95}. This parametrization is based on
vector-meson dominance with constraints imposed on the asymptotic
shape of $F_1$ and $F_2$ using arguments from perturbative QCD.

The only remaining freedom in the calculation is the choice of the
parameters governing the $\rho \pi \gamma$ MEC. The coupling $g_{\rho
  \pi \gamma}=0.563$ is extracted from the decay $\rho \rightarrow \pi
\gamma$. The contribution of the MEC to electron-deuteron scattering
depends crucially on the behavior of the current operator as a
function of $Q^2$.  In the work of Hummel and Tjon vector-meson
dominance was used to obtain a $\rho \pi \gamma$ form factor given
solely by the $\omega$ meson: $F_{\rho \pi \gamma}(Q^2)=\frac{1}{Q^2 -
  m_\omega^2}$.  This same $\rho \pi \gamma$ form factor is also
employed in the non-relativistic calculations of
Refs.~\cite{RS91,Wi95}. Other calculations have used form factors based
on quark models~\cite{vO95}. Such form factors tend to reduce the
contribution of this MEC, which is also very sensitive to the cutoff
masses in the $\pi NN$ and $\rho NN$ vertices.

\begin{figure}[h,t,b]
\centerline{\BoxedEPSF{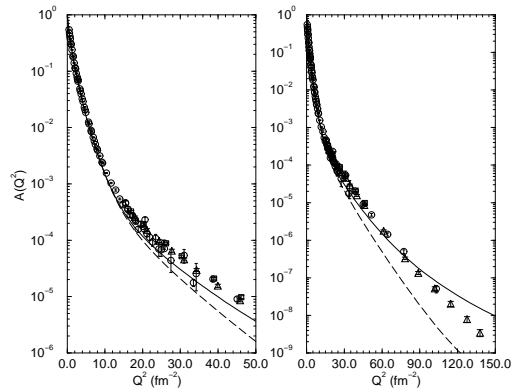 scaled 320}}

\caption{The deuteron structure function $A$. The left panel is an 
  enlarged version of the right one. In both the dashed line is the
  impulse approximation result, and the solid line is the result with
  the $\rho \pi \gamma$ MEC included. The experimental data of
  Refs.~[28,29] are denoted by the open circles,
  while that of Refs.~[26,27] are represented by
  triangles and squares.}
\label{fig-A}  
\end{figure}

In Fig.~\ref{fig-A} we present our results for $A$. The left panel
shows the results up to $Q^2 \approx 2$ GeV$^2$ and the right panel
gives results and data over the full range of experimental $Q^2$.  The
two JLab experiments which have produced data for
$A$~[Al98,Ab98] are denoted by triangles and squares
respectively. Note that these experiments confirm the trend of the
SLAC data of Arnold {\it et al.}

We see clearly in the left panel of Fig.~\ref{fig-A} that the impulse
approximation underpredicts the $A$ data for $Q^2=1-2$ GeV$^2$.  Much
of this lack of strength is made up for by the $\rho \pi \gamma$ MEC,
which gives a curve that reproduces the data in the region $Q^2=2-4 \,
{\rm GeV}^2$. However, this MEC calculation then {\it over}predicts
the JLab data at large $Q^2$.  Our $\rho \pi \gamma$ MEC contribution
to $A$ scales as $Q^{-20}$ when $Q$ is large, which is consistent
with perturbative QCD.  The failure of our calculation to describe the
data for $Q^2 > 4$ GeV$^2$ implies that the $\rho \pi \gamma$
mechanism is too strong at large $Q$ in the present model.

\begin{figure}[h,t,b]
\centerline{\BoxedEPSF{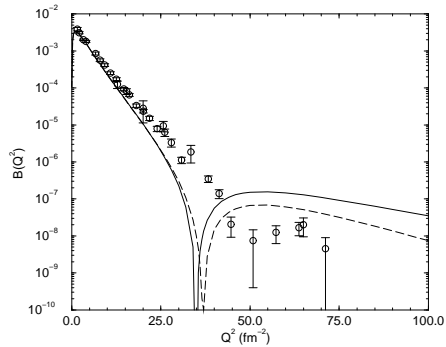 scaled 320}}

\caption{The deuteron structure function $B$. The experimental
  data are given by the open circles. The dashed and solid lines
  represent impulse approximation and $\rho \pi \gamma$ MEC
  calculations.}
\label{fig-B}  
\end{figure}

As seen in Fig.~\ref{fig-B} the extant experimental data for $B$
(which, as yet, includes no JLab data)~\cite{AandB,Bdata} are even less
well-described. Already at $Q^2 < 1$ GeV$^2$ there is significant
disagreement between our calculation and the data. This disagreement
is worsened by the inclusion of the relativistic version of the $\rho
\pi \gamma$ MEC: an effect also seen in the calculations of
Ref.~\cite{HT}.  There is a significant variation in results for this
observable at $Q^2=1-2$ GeV$^2$ in non-relativistic formulations of
this problem, partly because in this regime $B$ is sensitive to
currents associated with the short-range piece of the $NN$
interaction.  These currents, as well as other mechanisms not in our
model, could play a key role in reproducing the $B$ data.

\begin{figure}[h,t,b]
\centerline{\BoxedEPSF{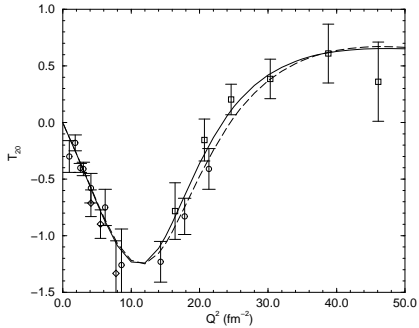 scaled 320}}

\caption{The tensor-polarization observable $T_{20}$. The older experimental
  data are denoted by the open circles, the recent NIKHEF data of
  Bouwhuis {\it et al.} by diamonds, and the JLab data of Beise {\it et al.} by
  squares. The dashed and solid line represent impulse approximation
  and $\rho \pi \gamma$ MEC calculations.}
\label{fig-T20}
\end{figure}

On the other hand, Fig.~\ref{fig-T20} shows that the $T_{20}$ data of
Refs.~\cite{T20data,largeQT20} is well-described by our approach out
to $Q^2 \approx 2 \, {\rm GeV}^2$.  This observable is fairly
insensitive to some of the dynamics which plays a role in $A$ and $B$
(e.g. the single-nucleon form factors). However, it is quite sensitive
to relativistic effects (see, for instance~\cite{Wi95,CK99}), so it is
gratifying that our approach reproduces the data, especially that of
Ref.~\cite{largeQT20} at large $Q^2$, so well.

\acknowledgements{We thank E.~J.~Beise and M.~Petratos for useful
  conversations. We are also grateful to the U.~S. Department of
  Energy, Nuclear Physics Division, for its support under grants
  DE-FG02-93ER-40762 and DE-FG03-97ER41014.}


\end{document}